%(April 6th revision)
\input harvmac.tex
\input epsf.tex
\parindent=0pt
\parskip=5pt

\hyphenation{satisfying}

\def\IR{{\hbox{{\rm I}\kern-.2em\hbox{\rm R}}}}
\def\IB{{\hbox{{\rm I}\kern-.2em\hbox{\rm B}}}}
\def\IN{{\hbox{{\rm I}\kern-.2em\hbox{\rm N}}}}
\def\IC{\,\,{\hbox{{\rm I}\kern-.59em\hbox{\bf C}}}}
\def\IZ{{\hbox{{\rm Z}\kern-.4em\hbox{\rm Z}}}}
\def\IP{{\hbox{{\rm I}\kern-.2em\hbox{\rm P}}}}
\def\IH{{\hbox{{\rm I}\kern-.4em\hbox{\rm H}}}}
\def\ID{{\hbox{{\rm I}\kern-.2em\hbox{\rm D}}}}

\def\N{{\cal N}}

\def\tr{{\rm tr}}

\noblackbox

\Title{\vbox{\baselineskip12pt
\hbox{UK/97--05}
\hbox{BUHEP--97--11}
\hbox{hep-th/9703210}}}
{Orientifolds, Branes, and Duality of 4D Gauge Theories }

\centerline{ \bf Nick Evans$^a$, Clifford V. Johnson$^b$ and 
Alfred D. Shapere$^c$ }
\bigskip\centerline{$^a${\it Department of Physics, Boston 
University, Boston, MA 02215 USA }}
\bigskip\centerline{$^{b,c}${\it Department of Physics and 
Astronomy, University of Kentucky, 
Lexington, KY 40506 USA}}
\footnote{}{\sl email: $^a${\tt nevans@physics.bu.edu}, 
$^b${\tt cvj@pa.uky.edu},  
$^c${\tt shapere@pa.uky.edu}}
\vskip1.5cm
\centerline{\bf Abstract}
\vskip0.7cm
\vbox{\narrower\baselineskip=12pt\noindent
Recently, a D--brane construction in type IIA string theory was shown
to yield the electric/magnetic duality of four dimensional $\N{=}1$
supersymmetric $U(N_c)$ gauge theories with $N_f$ flavours of quark.
We present here an extension of that construction which yields the
electric/magnetic duality for the $SO(N_c)$ and $USp(N_c)$ gauge
theories with $N_f$ quarks, by adding an orientifold plane which is
consistent with the supersymmetry.  Due to the intersection of the
orientifold plane with the NS--NS fivebranes already present, new
features arise which are crucial in determining the correct final
structure of the dualities.}
\vskip0.5cm

%\draft
\Date{March 1997}
%\Date{5th March 1997, My birthday! -cvj}

\baselineskip13pt

\lref\dbranes{J.~Dai, R.~G.~Leigh and J.~Polchinski,  Mod.~Phys.~Lett.
{\bf A4} (1989) 2073\semi P.~Ho\u{r}ava, Phys. Lett. {\bf B231} (1989)
251\semi R.~G.~Leigh, Mod.~Phys.~Lett. {\bf A4} (1989) 2767\semi
J.~Polchinski, Phys.~Rev.~D50 (1994) 6041, hep-th/9407031.}

\lref\orientifolds{A. Sagnotti, in {\sl `Non--Perturbative Quantum
 Field Theory'}, Eds. G. Mack {\it et. al.} (Pergammon Press, 1988), p521\semi
V. Periwal, unpublished\semi J. Govaerts, Phys. Lett. {\bf B220}
(1989) 77\semi P. Hor\u{a}va, Nucl. Phys. {\bf B327} (1989) 461.}

\lref\nsfivebrane{A. Strominger, Nucl. Phys. {\bf B343}, (1990) 167, 
{\it erratum} {\bf 353} (1991) 565.}
\lref\nscft{C. G. Callan, J.A. Harvey and A. Strominger, Nulc. Phys. 
{\bf B359} (1991), 611; Nucl. Phys. {\bf 367} (1991), 60.}
\lref\joetasi{J. Polchinski, 
hep-th/9611050.}

\lref\seiberg{N. Seiberg, Nucl. Phys. {\bf B435}, (1995) 129; hep-th/9411149.}
\lref\intriligatorthree{K. Intriligator and N. Seiberg, Phys. Lett.
 {\bf B387} (1996) 513; hep-th/9607207.}
\lref\intriligatorfour{K. Intriligator and N. Seiberg, Nucl. Phys. 
{\bf B444} (1995) 125; hep-th/9503179.}
\lref\intriligatorfive{R. G. Leigh and M. Strassler, Nucl. Phys.
 {\bf B356} (1995) 492; hep-th/9505088.}

\lref\hanany{A. Hanany and E. Witten,  hep-th/9611230.}

\lref\elitzur{S. Elitzur, A. Giveon and D. Kutasov,  hep-th/9702014.}

\lref\duffreport{M. J. Duff, J. X. Lu and  R. R. Khuri, Phys. Rept. 
{\bf 259} (1995) 213; hep-th/9412184.}

\lref\barbon{J. Barb\'{o}n,
 hep-th/9703051.}

\lref\BL{A. Brandhuber and K. Landsteiner, Phys.\ Lett.\ {\bf B358} (1995) 
73, hep-th/9507008.}

\lref\AS{P. Argyres and A. Shapere, Nucl.\ Phys.\ {\bf B461} (1996) 437, 
hep-th/9509175.}

\lref\wittennew{E. Witten,  hep-th/9703166.}

\lref\seibergwitten{N. Seiberg and E. Witten, 
Nucl.\ Phys.\ {\bf B426} (1994) 19, hep-th/9407087;
Nucl.\ Phys.\ {\bf B431} (1994) 484, hep-th/9408099.}

\lref\screen{T. Banks and E. Rabinovici, Nucl. Phys. {\bf B160} (1979) 
349\semi
E. Fradkin and S. Shenker, Phys. Rev. {\bf D19} (1979) 3682.}

\lref\APSei{P.C. Argyres, M.R. Plesser, and N. Seiberg, 
Nucl.\ Phys.\ {\bf B471} (1996) 159, hep-th/9603042.}

\lref\APS{P.C. Argyres, M.R. Plesser, and A.D. Shapere,
 Nucl.\ Phys.\ {\bf B483} (1997) 172,
hep-th/9608129.}

\lref\DS{U.H. Danielsson and B. Sundborg, 
 Phys.\ Lett.\ {\bf B358} (1995) 273, hep-th/9504102.}

\lref\petr{P. Hor\u{a}va, Phys. Lett. {\bf B289} (1991) 293; hep-th/9203031.}
\lref\moore{M. R. Douglas and G. Moore,  
 hep-th/9603167.}
\lref\ericjoe{E. G. Gimon and J. Polchinski, Phys. Rev. {\bf D54} (1996) 
1667, hep-th/9601038.}
\lref\ericmeI{E. G. Gimon and C. V. Johnson, Nucl. 
Phys. {\bf B477} (1996) 715, hep-th/9604129.}
\lref\joetensor{J. Polchinski,  
hep-th/9606165, to appear in Phys. Rev. {\bf D}.}
\lref\robme{C. V. Johnson and R. C. Myers, 
 hep-th/9610140, to appear in Phys. Rev. {\bf D}.}

\lref\orbifold{L. Dixon, J. Harvey, C. Vafa and E. Witten, 
 Nucl. Phys. {\bf B261} (1985) 678;
 Nucl. Phys. {\bf B274} (1986) 285.}
\lref\ooguri{H. Ooguri and C. Vafa, hep-th/9702180.}
%%%%%%%%%%%%%%%%%%%%%%%%%%%%%%%%%%%%%%%%%%%%%%%%%%%%%%%%%%%%%%%%%%%%%%%%%%

\newsec{Introduction}

\subsec{\sl Motivation}
Recently, a qualitatively new approach to applying string theory to
the study of field theory dualities has emerged. The field theories
are realised as limits of string vacua which are constructed as
configurations of intersecting (Ramond)$^2$ (R-R) branes and
(Neveu-Schwarz)$^2$ (NS--NS) branes in flat ten dimensional
spacetime. There is no involvement of non--trivial background fields
representing a curved background compact geometry. The structure of
the models is supplied purely by the intrinsic complexity of the brane
configurations themselves.

In ref.\hanany, a type~IIB string theory realisation of the $\N{=}4$
three dimensional `mirror' dualities of ref.\intriligatorthree\ was
presented. It employed an intricate interplay between an NS--NS
fivebrane\nsfivebrane, a D5--brane and a family of parallel
D3--branes\foot{We refer the reader to the literature for explanations
of the term `D-brane'\dbranes\ and `orientifold'\orientifolds, (to
appear later). For the definitive review (to date) see
ref.\joetasi. See also ref.\duffreport\ for a fine review of other
string soliton techniques.}. The world volumes of the branes were all
flat and fully extended, except for one dimension of the D3--branes,
which was a finite interval whose length was set by the distance
between the fivebranes. The $\N{=}4$ field theory was realised on the
(infinite part of) the world volume of the D3--branes.

The mirror duality was implemented by an exchange of the two species
of fivebrane (together with a rotation in some of the coordinates).  A
crucial ingredient was knowledge of the result of moving an NS--NS
fivebrane past a D5--brane. It is not a straightforward matter to
deduce the result of such a motion directly from string theory, as
this requires more knowledge about the description of NS--NS string
solitons than is presently available. This is largely because the
string coupling diverges at their core, taking us out of the regime
where we can presently directly calculate. However, continuity of the
BPS spectrum led the authors of ref.\hanany\ to realise that after
moving an NS--NS fivebrane past a D5--brane a new D3--brane must
appear stretched between them. This new feature was essential in
reconstructing the final dual theory.

In four dimensions, there are a number of situations where the
infrared (IR) limits of certain $\N{=}1$ supersymmetric gauge theories
have dual descriptions. In the case\seiberg\ of gauge group $U(N_c)$
with $N_f$ flavours of quark, (denoted $Q^i,{\tilde Q}^j\,\,
i,j=1,\ldots, N_f$) in the fundamental representation (the `electric'
scenario), the IR limit of the theory has a dual description in terms
of a $U(N_f-N_c)$ gauge theory with $N_f$ quarks ($q^i$, ${\tilde
q}^j$) in the fundamental (the `magnetic' scenario), together with a
$N_f^2$--component gauge singlet meson, $M_{ij}$. There is also a
coupling in the superpotential of the form $M_{ij}Q^i\cdot{\tilde Q}^{
j}$.

For the case\intriligatorfour\ of gauge group $SO(N_c)$ (or
$USp(N_c)$) with $N_f$ flavours of quark, there is also a dual
description of the IR regime, this time in terms of an $SO(N_f{-}N_c{+}4)$
(or $USp(N_f{-}N_c{-}4)$) gauge theory with $N_f$ flavours of quark.
There is again a meson, a symmetric tensor under the global flavour
symmetry for $SO$, (or an antisymmetric tensor for $USp$), and a
superpotential $M_{ij} q^i {\tilde q}^j$.

In ref.\elitzur, the duality for the $U(N_c)$ models was described 
using a type~IIA  string theory 
configuration inspired by ref.\hanany. This time, the ingredients 
were two   NS--NS fivebranes, arranged differently  in the ten 
dimensional space, a family of $N_c$ D4--branes, and a family of 
$N_f$ D6--branes. One of the dimensions of the world volume of the 
D4--branes was a finite segment stretched between the fivebranes. The 
field theory of interest lives on the infinite part of the world 
volume of the D4--branes.

The important feature here was inherited from the discussion of
ref.\hanany: When NS--NS fivebranes move past D6--branes, there is a
new D4--brane stretched between them. This conjectured behaviour was
exploited later in the construction to obtain the correct dual
theory. In addition, another type of unfamiliar strong coupling
behaviour could potentially have arisen in that paper: The physics of
two NS--NS fivebranes (with D4--branes connected to them) passing
through one another. This type of situation was avoided by moving one
fivebrane {\sl around} the other, using the freedom to move in
transverse directions.

In this paper we present a description of the $SO/USp$ dualities in
the spirit of refs.\refs{\hanany,\elitzur}\foot{Our approach is
complementary to that of ref.\ooguri. There, a relationship between
NS--NS fivebranes and certain singularities of Calabi--Yau manifolds
is used to rephrase the results of refs.\refs{\hanany,\elitzur}\ in a
geometrical context. That paper then presents a discussion of the
introduction of an orientifold into the geometrical framework in order
to derive the $SO/USp$ dualities. (We thank H.~Ooguri and C.~Vafa for
pointing out ref.\ooguri\ to us after the appearance of an earlier
version of this manuscript.)}. To do this we add a new ingredient, the
orientifold. As is by now well known (see ref.\joetasi\ for a review),
the orientifold is an extremely natural way of introducing orthogonal
and symplectic gauge groups into type~II string theory vacua, the
type~I string theory itself being the prototype example, with gauge
group $SO(32)$.

Most simply put, the orientifolding procedure combines the gauging of
fundamental string worldsheet parity $\Omega$ with target spacetime
discrete symmetries, resulting in the introduction of non--orientable
string sectors into the theory. The fixed points (in spacetime) of the
discrete symmetries are called `orientifold planes' (or sometimes just
`orientifolds') and can have any dimensionality. We shall call
orientifolds with a $(p{+}1)$~dimensional fixed plane `O$p$--planes',
in analogy with the term `D$p$--branes' we have been using for a
$(p{+}1)$~dimensional object.

There are a number of similarities and differences between O--planes
and D--branes. They are both extended objects. They both typically
break half of the supersymmetries. They both couple naturally to the
R--R sector fields in the theory. However, while the D-branes are
dynamical objects, the O--planes are not, at least in perturbation
theory.

There has been much work over the last year in studying string vacua
containing both orientifolds and D--branes. (Indeed, in many of those
cases, due to a compact transverse geometry the presence of one tends
to demand the presence of the other in order to satisfy
 one--loop consistency.)
There has been less work done in the context of mixing O--planes and
NS--NS branes. This is largely because of the lack of a complete
description of the latter type of object, a situation not unrelated to
the aforementioned strong coupling region at the core.

So although it is natural to introduce O--planes to facilitate the
description of the $SO$ and $USp$ dualities, we will inevitably have
to consider new phenomena.  On one hand, this is fortunate, as without
new phenomena we will not be able to describe the details of the
duality correctly. (For example, there will be a necessity for a pair
of D4--branes to appear (or disappear) as we move between dual
descriptions. Furthermore, at least for $SO(N_c)$ gauge theories,
there is a genuine phase transition, not present for the $SU(N_c)$
theories\seiberg\screen, which must also occur.) On the other hand, we
shall have to make some guesses about new phenomena in strongly
coupled string theory, as the orientifold forces us to consider new
strong coupling regions of the scenarios we construct.  On a third
hand, we can find strong justification of our new stringy results by
appealing to certain known properties of $\N{=}2$ theories which are
highly suggestive of non--perturbative physics attributable to
M--theory. So given that these new phenomena yield the $\N{=}1$ field
theory duality results we are studying, we can be very satisfied that
this enterprise has taught us some new details about a thorny problem
in strongly coupled string theory.

In the remainder of this introduction we recall some of the details of
the construction used in ref.\elitzur\ to realize electric and
magnetic $U(N_c)$ theories and the duality between them, within
type~IIA string theory. In section 2, we show how to do the same for
$SO(N_c)$ and $USp(N_c)$ theories by introducing an orientifold. Here
it will be necessary to make an assumption about the behaviour of
orientifolded NS--NS fivebranes at strong coupling. Evidence for our
assumption will be found in section 3, where we present a similar
brane realization of $\N{=}2$ gauge theories.  Specifically, we will
find that the auxiliary Riemann surface of Seiberg
and Witten\seibergwitten\ arises naturally, allowing us 
to directly identify the moduli of the brane configurations with
parameters of the effective field theory.  This correspondence is
interesting in its own right, and has been recently  independently
studied by Witten \wittennew\ in the context of $U(N_c)$ gauge
theories.  While our paper  represents work that was completed
before ref.\wittennew\ appeared, we expect that Witten's work will
play an important role in future developments.

In Section 4, we give a generalization of the construction of Section
2, which may describe theories with adjoint matter.  Finally, we
summarize our conclusions in Section 5.

\bigskip
\bigskip
\bigskip

\subsec{\sl The `electric' $U(N_c)$ gauge theory.}

The brane configuration of ref.\elitzur\ may be summarized by the
following table\foot{The temptation to term such a table a
`brane--scan' is nearly overwhelming.}:

%\topinsert{
\bigskip
\vbox{
$$\vbox{\offinterlineskip
\hrule height 1.1pt
\halign{&\vrule width 1.1pt#
&\strut\quad#\hfil\quad&
\vrule#
&\strut\quad#\hfil\quad&
\vrule width 1.1pt#
&\strut\quad#\hfil\quad&
\vrule#
&\strut\quad#\hfil\quad&
\vrule#
&\strut\quad#\hfil\quad&
\vrule#
&\strut\quad#\hfil\quad&
\vrule#
&\strut\quad#\hfil\quad&
\vrule#
&\strut\quad#\hfil\quad&
\vrule#
&\strut\quad#\hfil\quad&
\vrule#
&\strut\quad#\hfil\quad&
\vrule#
&\strut\quad#\hfil\quad&
\vrule#
&\strut\quad#\hfil\quad&
\vrule width 1.1pt#\cr
height3pt
&\omit&
&\omit&
&\omit&
&\omit&
&\omit&
&\omit&
&\omit&
&\omit&
&\omit&
&\omit&
&\omit&
&\omit&
\cr
&\hfil type&
&\hfil \#&
&\hfil $x^0$&
&\hfil $x^1$&
&\hfil $x^2$&
&\hfil $x^3$&
&\hfil $x^4$&
&\hfil $x^5$&
&\hfil $x^6$&
&\hfil $x^7$&
&\hfil $x^8$&
&\hfil $x^9$&
\cr
height3pt
&\omit&
&\omit&
&\omit&
&\omit&
&\omit&
&\omit&
&\omit&
&\omit&
&\omit&
&\omit&
&\omit&
&\omit&
\cr
\noalign{\hrule height 1.1pt}
height3pt
&\omit&
&\omit&
&\omit&
&\omit&
&\omit&
&\omit&
&\omit&
&\omit&
&\omit&
&\omit&
&\omit&
&\omit&
\cr
&\hfil NS&
&\hfil $1$&
&\hfil --- &
&\hfil --- &
&\hfil --- &
&\hfil --- &
&\hfil --- &
&\hfil --- &
&\hfil $\bullet$ &
&\hfil $\bullet$ &
&\hfil $\bullet$ &
&\hfil $\bullet$ &
\cr
height3pt
&\omit&
&\omit&
&\omit&
&\omit&
&\omit&
&\omit&
&\omit&
&\omit&
&\omit&
&\omit&
&\omit&
&\omit&
\cr
\noalign{\hrule}
height3pt
&\omit&
&\omit&
&\omit&
&\omit&
&\omit&
&\omit&
&\omit&
&\omit&
&\omit&
&\omit&
&\omit&
&\omit&
\cr
&\hfil NS$^\prime$&
&\hfil $1$&
&\hfil --- &
&\hfil --- &
&\hfil --- &
&\hfil --- &
&\hfil $\bullet$ &
&\hfil $\bullet$ &
&\hfil $\bullet$ &
&\hfil $\bullet$ &
&\hfil --- &
&\hfil --- &
\cr
height3pt
&\omit&
&\omit&
&\omit&
&\omit&
&\omit&
&\omit&
&\omit&
&\omit&
&\omit&
&\omit&
&\omit&
&\omit&
\cr
\noalign{\hrule}
height3pt
&\omit&
&\omit&
&\omit&
&\omit&
&\omit&
&\omit&
&\omit&
&\omit&
&\omit&
&\omit&
&\omit&
&\omit&
\cr
&\hfil D4&
&\hfil $N_c$&
&\hfil --- &
&\hfil --- &
&\hfil --- &
&\hfil --- &
&\hfil $\bullet$ &
&\hfil $\bullet$ &
&\hfil [---] &
&\hfil $\bullet$ &
&\hfil $\bullet$ &
&\hfil $\bullet$ &
\cr
height3pt
&\omit&
&\omit&
&\omit&
&\omit&
&\omit&
&\omit&
&\omit&
&\omit&
&\omit&
&\omit&
&\omit&
&\omit&
\cr
\noalign{\hrule}
height3pt
&\omit&
&\omit&
&\omit&
&\omit&
&\omit&
&\omit&
&\omit&
&\omit&
&\omit&
&\omit&
&\omit&
&\omit&
\cr
&\hfil D6&
&\hfil $N_f$&
&\hfil --- &
&\hfil --- &
&\hfil --- &
&\hfil --- &
&\hfil $\bullet$ &
&\hfil $\bullet$ &
&\hfil $\bullet$ &
&\hfil --- &
&\hfil --- &
&\hfil --- &
\cr
height3pt
&\omit&
&\omit&
&\omit&
&\omit&
&\omit&
&\omit&
&\omit&
&\omit&
&\omit&
&\omit&
&\omit&
&\omit&
\cr
}\hrule height 1.1pt
}
$$
}
\centerline{\sl Table 1.}

\bigskip
In the table, a dash `---' represents a direction {\sl along} a
brane's worldvolume while a dot~`$\bullet$' is transverse. For the
special case of the D4--branes' $x^6$ direction, where a worldvolume
is a finite interval, we use the symbol `{\rm [---]}'.  (It is
particularly simple to read off a lot of information from such a
table. For example a `$\bullet$' and a `---' in the same column says
that one object is living inside the worldvolume of the other in that
direction, and so they can't avoid one another. Meanwhile two
`$\bullet$'s in the same column tell us that the objects are
pointlike, and need not coincide in that direction, except for the
specific case where they share identical values of that coordinate.)

The $N_c$ coincident D4--branes give rise to a $U(N_c)$ gauge symmetry
on their worldvolumes. This symmetry arises from massless fundamental
strings (`4--4 strings') connecting the various branes, in the usual
way. Focusing on the four dimensions of the $(x^0,x^1,x^2,x^3)$
directions, we have a gauge theory with coupling strength given by
$g^2\sim 1/L_6$, where $L_6$ is the distance the D4--branes are
stretched between the two fivebranes in the $x^6$ direction.

The $N_f$ (not necessarily coincident) D6--branes contribute (via 6--4
strings) matter fields to the $U(N_c)$ gauge theory, transforming in
the $N_c$ dimensional fundamental representation. There are $N_f$
flavours of such quarks.

There are also 6--6 strings, whose role is to supply a flavour
symmetry to the problem which is generically $U(1)^{N_f}$, but can be
as large as $U(N_f)$.  As this symmetry is a gauge symmetry on the
seven dimensional world volumes of the D6--branes, it is best thought
of as a global `spectator' symmetry from the point of view of the
dynamics of the D4--branes' worldvolume gauge theory. It will not play
a major role in the proceedings.

That it is $\N{=}1$ supersymmetry which is present in four dimensions
follows from analyzing the conditions on the spinor generators imposed
by the worldvolumes of the various objects.  The analysis is already
presented in ref.\elitzur, generalising the presentation in
ref.\hanany, and will not undergo any modification here. The
configuration preserves 1/16 of the original ten dimensional $\N{=}2$
supersymmetry.

For definiteness, take the configuration giving this `electric' type
description of the field theory to have the first NS--NS fivebrane
(denoted NS), to the left of the second NS--NS fivebrane (denoted
NS$^\prime$) in the $x^6$ direction. There are $N_c$ D4--branes
stretched between them, along that direction, passing $N_f$ D6--branes
along the way.  (See Fig.~1.\foot{Note that in this and all other figures,
we have displaced the D4--branes away from coincidence, to aid with
visualisation, and we have ignored the $(x^0,x^1,x^2,x^3)$ directions 
which are common to all of the branes. Also, we will only indicate on a 
diagram  whether the branes are pointlike or extended in the
 $(x^4,x^5,x^7,x^8,x^9)$ directions when necessary. 
That information may be found in Tables~1 and~2.})
 
\midinsert{
\vskip0.1cm
\hskip3.1cm
\epsfxsize=3.0truein\epsfbox{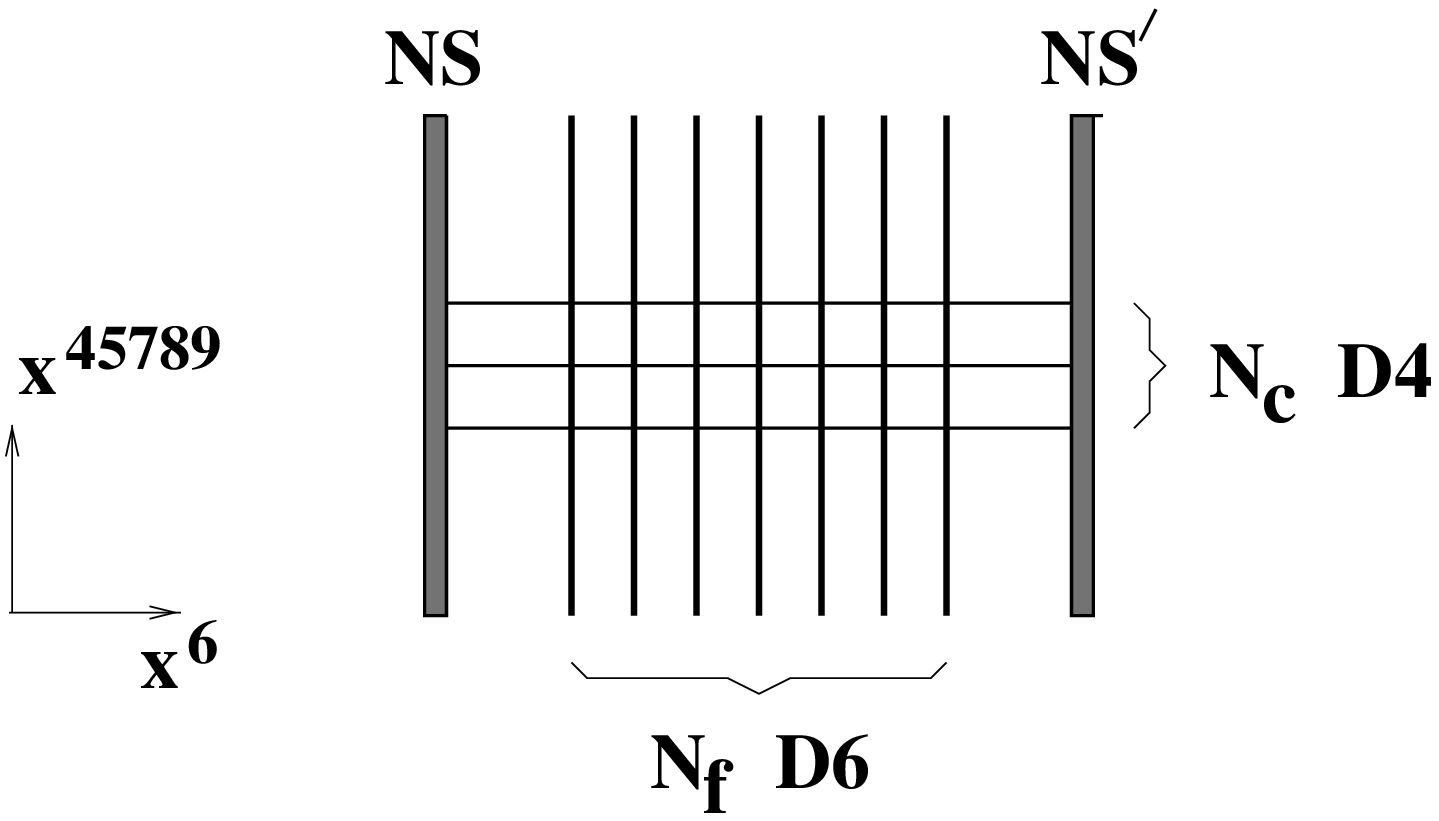}
\vskip0.1cm
\centerline{\sl Figure 1.}
}\endinsert

Note that the translational degrees of freedom of the string model
have a one--to--one mapping to moduli of the gauge theory. The
distances between the D6--branes and the D4--branes in the $(x^4,x^5)$
direction correspond to mass terms for the matter fields. Segments of
D4--branes connecting to the NS$^\prime$ fivebrane and touching a
D6--brane, and those stretching between D6--branes, are free to move
in the $(x^8,x^9)$ direction.  In general, such movement will leave
the $N_c$ D4--branes non--coincident, and therefore these directions
correspond to vacuum expectation values (vev's) for the matter fields
that break the gauge symmetry.

The final translation that may be made is to move the two NS--NS
fivebranes relative to each other in the $x^7$ direction. This may be
done if we also shift the $x^7$ positions where the $N_c$ D4--branes
touch the D6--branes. This potentially breaks supersymmetry by
introducing a Fayet--Iliopoulos (FI) term into the potential of the
field theory. With massless matter fields present, supersymmetry
breaking may be avoided by turning on matter field vev's that generate
a non--zero $D$--term (canceling the FI term) and break the gauge
symmetry. Such FI terms can only arise in theories which have a $U(1)$
centre of the gauge group\foot{It is easy to see on general grounds
that this coupling must enter the field theory in this way. So far,
everything else in the effective field theory are fields from the open
string sector, coming in vector--and hyper--multiplets of the gauge
and supersymmetry. The positions of the NS--NS fivebranes are governed
by the closed string sector, contributing a new type of term to the
Lagrangian which does not transform under the gauge symmetry of the
open string sector.  It enters as a gauge invariant
supersymmetry--breaking term. The possibility of restoring
supersymmetry by a Higgs mechanism allows for a new direction in the
moduli space of vacua, a Higgs phase. Simliar reasoning has been used
in other situations, for example in identifying the FI terms
corresponding to blowing up an ALE
space\refs{\moore,\joetensor,\robme}.}.

\subsec{\sl The `magnetic' $U(N_f{-}N_c)$ gauge theory.}

Continuing to follow ref.\elitzur, the dual description of this theory
is obtained by exchanging the positions of the fivebranes in the $x^6$
direction. In order for this to happen, the NS fivebrane has to first
move past the D6--branes, which are at a definite values of $x^6$.

This is where the observation of ref.\hanany\ is crucial.  A study of
the spectrum in the worldvolume gauge theory of a related situation
(NS--NS fivebranes with D3--branes stretched between them, passing
through D5--branes) showed that there must be a new stretched brane
between the NS--NS fivebrane and the D--brane it passed through. This
may be deduced by insisting that if the movement is a true modulus of
the theory (which it is, as we can see in the field theory since
changing the ultra--violet (UV) coupling does not effect the far IR
behaviour of the theory) then the BPS spectrum must be the same before
and after the encounter. In order that there be the same
hypermultiplet structure before and after, the most conservative
explanation is that there is a new stretched D3--brane (the new
hypermultiplet corresponds to strings connecting the new D3--brane and
the old D3--brane).  This surmounts the problem of trying to describe
directly the strong coupling string physics lurking at the core of the
fivebrane due to the growth of the dilaton there.

So in the present context, when the NS fivebrane has moved past all of
the D6--branes, there are $N_f$ new D4--branes stretched in the $x^6$
direction. In particular, there is one stretching from each of the
$N_f$ D6--branes to the NS fivebrane. There are still $N_c$ D4--branes
between the two fivebranes. (See Fig.~2.)

\midinsert{
\vskip0.1cm
\hskip3.0cm
\epsfxsize=3.5truein\epsfbox{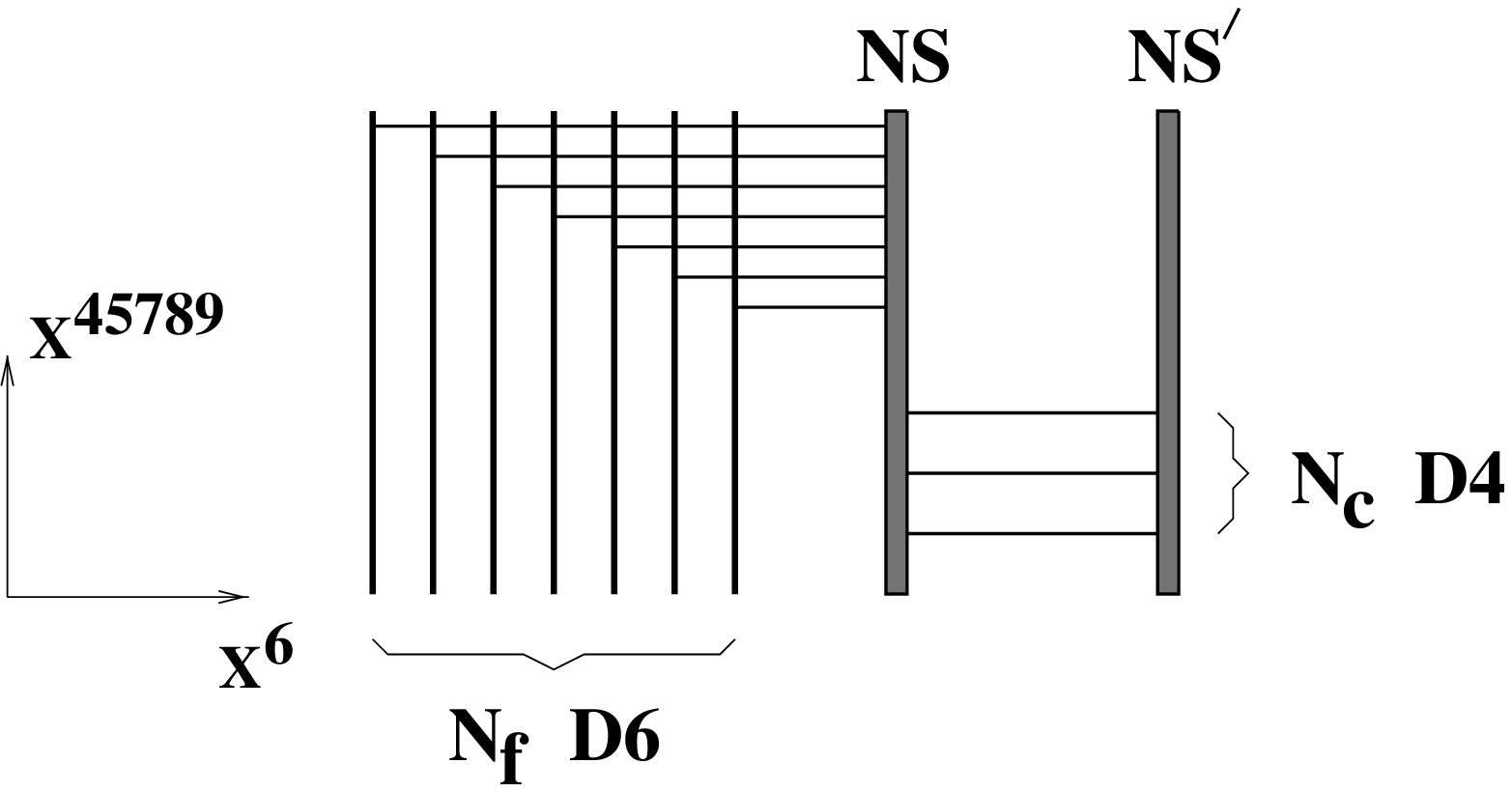}
\vskip0.1cm
\centerline{\sl Figure 2.}}\endinsert

The next step is to move past the NS$^\prime$ fivebrane.  In moving
the NS fivebrane --- with its entourage of D4--branes --- through the
NS$^\prime$ fivebrane, there is the aforementioned problem that we
have little knowledge of the description of string theory in such a
situation.  The presentations in refs.\refs{\hanany,\elitzur}
cunningly avoided this problem by going {\sl around} the potentially
singular behaviour.

There is the possibility to move the NS fivebrane off to a different
$x^7$ value than where the NS$^\prime$ fivebrane is located. It can
then go around and return to its original $x^7$ value once it has
moved far enough in the $x^6$ direction, thus ending up to the right
of the NS$^\prime$ fivebrane, achieving the desired final
configuration without encountering a new region of strong coupling.

As a result of the NS fivebrane moving off into the $x^7$ direction,
the $N_c$ connecting D4--branes can no longer stretch directly between
it and the NS$^\prime$ brane, as an examination of the configuration
table confirms. Instead, they connect from the NS$^\prime$ brane to
$N_c$ of the $N_f$ D6--branes (which are sharing the $(x^4, x^5)$
position of the NS$^\prime$ brane). The remaining $N_f{-}N_c$
D6--branes retain their connection to the NS brane. (See Fig.~3.)

\midinsert{
\vskip0.1cm
\hskip2.3cm
\epsfxsize=5.0truein\epsfbox{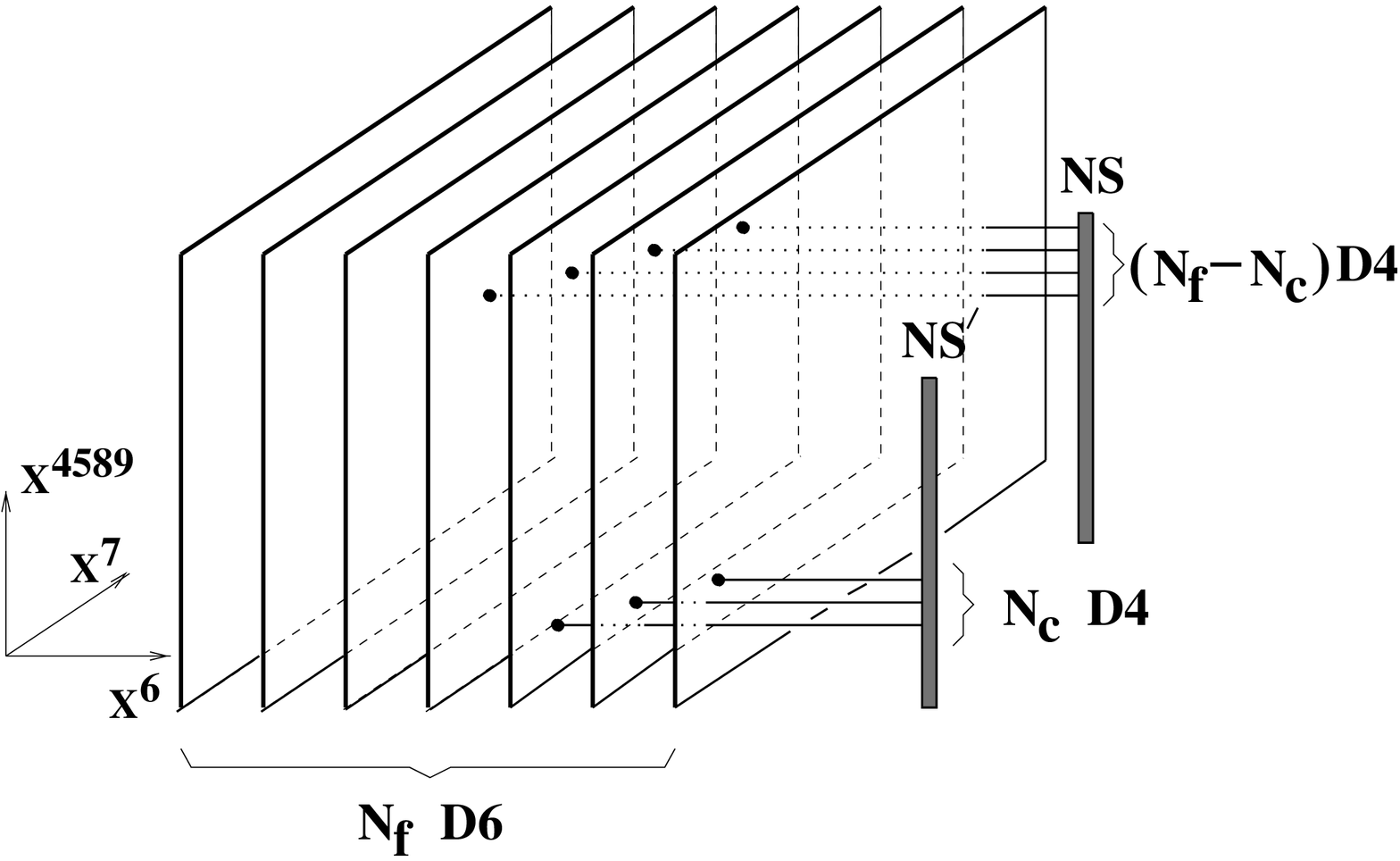}
\vskip0.1cm
\centerline{\sl Figure 3.}}\endinsert

The field theory is now in a Higgs phase. The distance between the
$x^7$ positions of the two fivebranes corresponds to an FI term in the
scalar potential of the theory.  In order to ensure supersymmetry, a
new zero of the scalar potential may be found by breaking the gauge
group with a Higgs mechanism, which is achieved by the just--described
movement of the D4--branes.

After moving around the NS$^\prime$ brane, the NS brane may return to
its original position in $x^7$.  The FI term disappears and a gauge
symmetry returns with the possibility of reconnecting the fivebranes
directly with D4--branes. The $N_f{-}N_c$ such branes connecting the
NS brane to the $N_f{-}N_c$ D6--branes now split, reconnecting free
ends to the NS$^\prime$ brane.  The $N_c$ D4--branes connecting the
D6--branes to the NS$^\prime$ brane are now accompanied by the
$N_f{-}N_c$ D4--branes from the other half of the split, now making
the same D6--NS$^\prime$ connection. (See Fig.~4.)

\midinsert{
\vskip0.1cm
\hskip2.0cm
\epsfxsize=4.0in\epsfbox{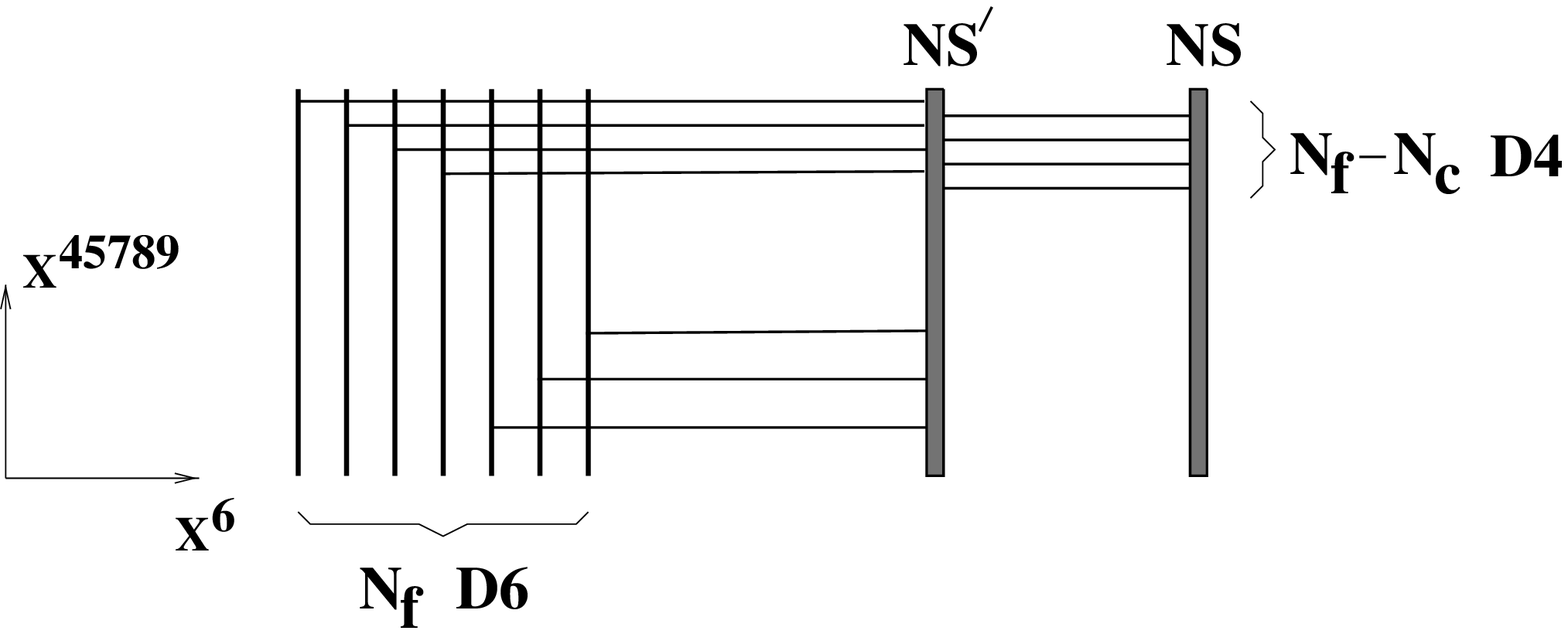}
\vskip0.1cm
\centerline{\sl Figure 4.}}
\endinsert

The worldvolume theory of the D4--branes is now as follows.  There is
a $U(N_f{-}N_c)$ gauge theory (from 4--4 strings between the fivebranes) 
with $N_f$ flavours
(from 4--4 strings across the NS$^\prime$ fivebrane) 
of quark, $q_i$, in the fundamental. There is also
a family of $N_f^2$ fields coming from the fluctuations (in ($x^8,
x^9$)) of 4--4 strings connecting the $N_f$ D4--branes. This is the
meson field $M_{ij}$. Its coupling to the quarks in the
superpotential may be deduced by examining the last figure and
considering how to turn 4--4 strings which define the quarks into 4--4
strings which make the meson.

This is the `magnetic' dual description of the original $U(N_c)$ field
theory, as first presented in ref.\seiberg.  The `loom' arrangement
above, weaving D4--branes between fivebranes and D6--branes, bears
fruit by making this $\N{=}1$ duality manifest, an (almost) simple
consequence of being embedded in string theory in this way.

\newsec{Electric/magnetic duality in $SO(N_c)$ and 
$USp(N_c)$ gauge theories.}

In this section we describe how to construct the electric/magnetic
duality for the $SO(N_c)$ and $USp(N_c)$ gauge theories with $N_f$
quarks. On general grounds, we expect that this will not be as
straightforward as the construction for the $U(N_c)$ gauge theories
(with $N_f$ quarks) for the following reason: There is no true phase
transition in going between the electric and magnetic descriptions for
the $SU(N_c)$ case, while there is a phase transition for 
$SO(N_c)$ \screen. The occurrence of a phase transition is
detected with an order parameter, which in this case is a Wilson loop
(in any representation).  In the case of $SU(N_c)$ with matter in the
fundamental, any phase transition which could be detected by such an
order parameter is `screened' by the quarks. This is because any
representation can be constructed with the fundamental (and
anti--fundamental) representations. In the case of $SO(N_c)$,
matter in the fundamental cannot screen sources which are in spinor
representations (which cannot be made out fundamentals) and so there
is a viable order parameter. A phase transition in going between the
electric and magnetic descriptions can be detected by observing that a
Wilson loop expectation value's dependence on the geometry 
changes from an area law to a perimeter law, or {\it vice versa}.

We do not expect, therefore, that we should be able to find a
description of the path between the magnetic and electric variables
which is as smooth as the one found in ref.\elitzur. We do expect that
at some point on the path between the two phases, a non--trivial point
{\sl must} be encountered which allows for the occurrence of a phase
transition. Indeed, we will find in this section that such a point is
{\sl forced upon us} by the presence of the orientifold which we
introduce in order to construct the $SO/USp(N_c)$ cases.

\subsec{\sl The new ingredient: An orientifold.}
Orthogonal and symplectic gauge  groups arise naturally in string  
theory  in the presence of D--branes by orientifolding. 

In general, the procedure of orientifolding will reduce the amount of
supersymmetry by half. This happens in much the same way as it happens
for D--branes. The worldvolume of an O--plane reflects the
supersymmetry generators, leaving only a linear combination of the
ingoing and outgoing spinor to carry the remaining supersymmetry. In
general, that will spoil our present arrangement considerably, as we
will fully break the four dimensional supersymmetry. It is possible,
however, in this situation to perform an orientifold in such a way as
to preserve the supersymmetry that is already present.

We need only orientifold in such a way as to create an O--plane whose
world--volume lies in the same dimensions as are already occupied by
the worldvolume of a D--brane. Then the O--plane's world volume will
place conditions on the spinors which are already satisfied, thus
preserving the supersymmetry of our arrangement.

After a little thought, it is obvious that we need only add an
O4--plane to our arrangement to get the desired result\foot{Of course,
we can also add an O6--plane parallel to the D6--branes and preserve
the same amount of supersymmetry. We will not do this here.}. This
O4--plane must extend in the $x^0,x^1,x^2, x^3, x^6$ directions to
preserve supersymmetry. Such a plane results from combining the
gauging of world sheet parity~$\Omega$ with the spacetime reflection
$$(x^4,x^5,x^7,x^8,x^9)\to (-x^4,-x^5,-x^7,-x^8,-x^9).$$ (Of course,
if these directions were compact, the resulting orbifolded torus would
have $32$ O4--planes, which is rather more than we need.)

Note that  the O4--plane is not of finite extent in the $x^6$
direction. Attaching it to the NS  and NS$^\prime$ fivebranes 
and moving it around with them would be tantamount to making it 
dynamical, which it is not, at least in string perturbation theory.

\subsec{\sl The `electric' $SO/USp(N_c)$ gauge theories.}

The configuration table for our new electric scenario is as follows:
%\topinsert{
\bigskip
\vbox{
$$\vbox{\offinterlineskip
\hrule height 1.1pt
\halign{&\vrule width 1.1pt#
&\strut\quad#\hfil\quad&
\vrule#
&\strut\quad#\hfil\quad&
\vrule width 1.1pt#
&\strut\quad#\hfil\quad&
\vrule#
&\strut\quad#\hfil\quad&
\vrule#
&\strut\quad#\hfil\quad&
\vrule#
&\strut\quad#\hfil\quad&
\vrule#
&\strut\quad#\hfil\quad&
\vrule#
&\strut\quad#\hfil\quad&
\vrule#
&\strut\quad#\hfil\quad&
\vrule#
&\strut\quad#\hfil\quad&
\vrule#
&\strut\quad#\hfil\quad&
\vrule#
&\strut\quad#\hfil\quad&
\vrule width 1.1pt#\cr
height3pt
&\omit&
&\omit&
&\omit&
&\omit&
&\omit&
&\omit&
&\omit&
&\omit&
&\omit&
&\omit&
&\omit&
&\omit&
\cr
&\hfil type&
&\hfil \#&
&\hfil $x^0$&
&\hfil $x^1$&
&\hfil $x^2$&
&\hfil $x^3$&
&\hfil $x^4$&
&\hfil $x^5$&
&\hfil $x^6$&
&\hfil $x^7$&
&\hfil $x^8$&
&\hfil $x^9$&
\cr
height3pt
&\omit&
&\omit&
&\omit&
&\omit&
&\omit&
&\omit&
&\omit&
&\omit&
&\omit&
&\omit&
&\omit&
&\omit&
\cr
\noalign{\hrule height 1.1pt}
height3pt
&\omit&
&\omit&
&\omit&
&\omit&
&\omit&
&\omit&
&\omit&
&\omit&
&\omit&
&\omit&
&\omit&
&\omit&
\cr
&\hfil NS&
&\hfil $1\over2$&
&\hfil --- &
&\hfil --- &
&\hfil --- &
&\hfil --- &
&\hfil --- &
&\hfil --- &
&\hfil $\bullet$ &
&\hfil $\bullet$ &
&\hfil $\bullet$ &
&\hfil $\bullet$ &
\cr
height3pt
&\omit&
&\omit&
&\omit&
&\omit&
&\omit&
&\omit&
&\omit&
&\omit&
&\omit&
&\omit&
&\omit&
&\omit&
\cr
\noalign{\hrule}
height3pt
&\omit&
&\omit&
&\omit&
&\omit&
&\omit&
&\omit&
&\omit&
&\omit&
&\omit&
&\omit&
&\omit&
&\omit&
\cr
&\hfil NS$^\prime$&
&\hfil $1\over2$&
&\hfil --- &
&\hfil --- &
&\hfil --- &
&\hfil --- &
&\hfil $\bullet$ &
&\hfil $\bullet$ &
&\hfil $\bullet$ &
&\hfil $\bullet$ &
&\hfil --- &
&\hfil --- &
\cr
height3pt
&\omit&
&\omit&
&\omit&
&\omit&
&\omit&
&\omit&
&\omit&
&\omit&
&\omit&
&\omit&
&\omit&
&\omit&
\cr
\noalign{\hrule}
height3pt
&\omit&
&\omit&
&\omit&
&\omit&
&\omit&
&\omit&
&\omit&
&\omit&
&\omit&
&\omit&
&\omit&
&\omit&
\cr
&\hfil O4&
&\hfil $1$&
&\hfil --- &
&\hfil --- &
&\hfil --- &
&\hfil --- &
&\hfil $\bullet$ &
&\hfil $\bullet$ &
&\hfil --- &
&\hfil $\bullet$ &
&\hfil $\bullet$ &
&\hfil $\bullet$ &
\cr
height3pt
&\omit&
&\omit&
&\omit&
&\omit&
&\omit&
&\omit&
&\omit&
&\omit&
&\omit&
&\omit&
&\omit&
&\omit&
\cr
\noalign{\hrule}
height3pt
&\omit&
&\omit&
&\omit&
&\omit&
&\omit&
&\omit&
&\omit&
&\omit&
&\omit&
&\omit&
&\omit&
&\omit&
\cr
&\hfil D4&
&\hfil $N_c\over2$&
&\hfil --- &
&\hfil --- &
&\hfil --- &
&\hfil --- &
&\hfil $\bullet$ &
&\hfil $\bullet$ &
&\hfil [---] &
&\hfil $\bullet$ &
&\hfil $\bullet$ &
&\hfil $\bullet$ &
\cr
height3pt
&\omit&
&\omit&
&\omit&
&\omit&
&\omit&
&\omit&
&\omit&
&\omit&
&\omit&
&\omit&
&\omit&
&\omit&
\cr
\noalign{\hrule}
height3pt
&\omit&
&\omit&
&\omit&
&\omit&
&\omit&
&\omit&
&\omit&
&\omit&
&\omit&
&\omit&
&\omit&
&\omit&
\cr
&\hfil D6&
&\hfil $N_f\over2$&
&\hfil --- &
&\hfil --- &
&\hfil --- &
&\hfil --- &
&\hfil $\bullet$ &
&\hfil $\bullet$ &
&\hfil $\bullet$ &
&\hfil --- &
&\hfil --- &
&\hfil --- &
\cr
height3pt
&\omit&
&\omit&
&\omit&
&\omit&
&\omit&
&\omit&
&\omit&
&\omit&
&\omit&
&\omit&
&\omit&
&\omit&
\cr
}\hrule height 1.1pt
}
$$
}
\centerline{\sl Table 2.}

\bigskip

Some of the basic effects of the orientifold are easy to describe,
referring to the table. Consider the directions $x^m$ where the
orientifold plane is located at a point.  Any object which is not
coincident with it in those dimensions (say at $x^m{=}x^m_0$) will
have a mirror copy of itself at $x^m{=}{-}x^m_0$.  This is why we have
the factors of one half in the counting of the number of physical
objects in each row of the table. It would be overcounting to consider
an object and its reflection as separate physical objects.

We will take $N_c/2$ D4--branes with their duplicates. Generically,
the gauge group is then $U(1)^{N_c/2}$.  If they are all coincident,
it is $U(N_c/2)$. However, when they are all coincident and lying
precisely on the O4--plane, strings between the $N_c/2$ D4--branes and
their copies fill out gauge group $SO(N_c)$ or $USp(N_c)$.  Whether
the gauge group is $SO(N_c)$ or $USp(N_c)$ results from the choice of
whether $\Omega^2$ acts as $\pm1$ on the open string sectors\ericjoe.

Also correlated with the sign of $\Omega^2$ is the
$H^{(6)}{=}dA^{(5)}$ R-R charge of the orientifold plane. In the
natural normalisation where the D4--branes carry one unit of this
charge, the O4--plane carries $\mp1$ units, for $\Omega^2{=}\pm1$ in
the D4--brane sector.

(It is worth noting that odd $N_c$ is achievable by the introduction
of ``half--D4--branes'' that are forced to remain in the O4--plane. In
general though, without a more complicated scenario than we have here,
only even numbers of half branes can move off the orientifold
plane. This translates into a pattern of Higgsing (and giving mass
terms in the dual theory) which can only change $N_c$ by two. We can
thus only deform the theory by relating even $N_c$ theories or odd
$N_c$ theories, which is a subset of the possible deformations of the
theory. For definiteness, we consider only even $N_c$ theories, but
note that we can consider odd $N_c$ theories, with the mentioned
restrictions on the type of deformations we can do. It is possible
that there are other scenarios, which will yield the even--to--odd
$N_c$ deformations that we don't see here.)

For the branes which are not completely parallel to the O4--plane,
things are interesting. The O4--plane cuts through them, and reflects
the physics on one side of the bisection into that on the other
side. Differently put, the O4--plane places a reflecting boundary in
the (parts of) the worldvolumes of the branes it intersects.
Referring to the table, this happens for directions in which an object
has a `---' where the O4--plane has a `$\bullet$'.

For the D6--branes, this is an interesting but completely innocuous
situation from the point of view of computing in weak coupling string
theory, as the orientifold is simply an additional projection
condition over and above the Dirichlet boundary conditions which
describe the D6--brane. If the D6--branes are moved off the O4--plane,
multiple copies will be generated and it is then clear that $N_f/2$
D6--branes give rise to $N_f$ matter multiplets\foot{Consistency of
the string theory requires that the possible gauged flavour symmetry
group coming from the $N_f/2$ D6--branes be $USp(N_f)$ or $SO(N_f)$
for the choices $\Omega^2=\pm1$ in the D4--brane sector. This is
because $\Omega^2$ will act with the opposite sign in the D6--brane
sector, a fact that is $T_{45789}$--dual to the situation with D5--
and D9--branes in type~I string theory\ericjoe. This requirement on
the possible gauged flavour symmetry is known independently from a
field theory perspective\APS.}.  Again odd numbers of flavours may
be generated by the inclusion of half D6--branes fixed on the
O4--plane.

However, for the NS--NS fivebranes, the physics of orientifolding is
not as clear.  There is certainly a partial description of this
situation in terms of an orientifold of the conformal field
theory\nscft\ of (part of) these objects. Indeed, such a description
is almost certainly related to some of the earliest non--trivial
orientifolds, studied in the context of black hole physics in
ref.\petr. There, the action of~$\Omega$ was gauged in combination
with a target space symmetry of a gauged WZW model. Notice that the
`throat' conformal field theories of NS--NS fivebranes are realised as
closely related gauged WZW models.  It would certainly be interesting
to compute some of the details of such a new situation as an NS--NS
fivebrane straddling an orientifold plane. (This opens up a
potentially vast area of investigation: revisiting many non--trivial
conformal field theories representing type~II backgrounds and
orientifolding them.  However, we will leave that as a future
direction of research, and press on with the errand of this paper.)

Considering an orientifolded NS--NS fivebrane in isolation for a
moment, we can anticipate some of the principal players in the content
of the resulting model.  There will be a new family of closed string
fields arising from orientifolding the closed strings making up the
NS--NS fivebrane. These come from the twisted sectors of the orbifold
part of the spectrum. In general, we expect that the O4--plane and the
NS--NS fivebrane must carry `twisted sector charges' under these
closed string fields.  As twisted sector fields have no zero mode (and
are therefore localised), sources for them must remain trapped at the
orbifold fixed point\orbifold.  This must mean that a NS--NS fivebrane
must remain on the O4--plane\foot{See
refs.\refs{\ericmeI,\joetensor,\robme}\ for other situations of
exactly this type.}. In this sense, the NS--NS fivebranes are really
half--fivebranes, analogous to the half D--branes mentioned earlier.  

Another way to see that the half--fivebrane is trapped on the orientifold is
from the content of the field theory. The gauge group (either
$SO(N_c)$ or $USp(N_c)$) now has no $U(1)$ center. Therefore any
coupling arising in the theory corresponding to moving the half--fivebrane
off into the $x^7$ direction cannot enter as a Fayet--Iliopoulos
term, and therefore it is impossible to arrange to Higgs the gauge
group in such as way as to make that type of movement a supersymmetric
flat direction of the theory.

This would seem to place a spanner in the works of our weaving
arrangement. Now, we cannot move the NS half--fivebrane off the orientifold
in the $x^7$ direction, as we wish to do to mimic the constructions of
refs.\refs{\hanany,\elitzur}.

We therefore cannot avoid the strong coupling singularity of moving
the NS--NS fivebranes through each other. Recall that we anticipated
the necessity of encountering such a singular situation at the
beginning of this section.  Without such a new feature, it is
difficult to see how the phase transition in moving between the
magnetic and electric phases of the $SO$ theories could
occur. Given that this is so, we are still confronted with the fact
that we have no accurate string theory description of what occurs at
such a point.  Undaunted, let us proceed to try to reach the magnetic
theories.

\subsec{\sl The `magnetic' $SO/USp(N_f{-}N_c\pm4)$ gauge theories.}

We begin with $N_c/2$ D4--branes stretched between the NS and the
NS$^\prime$ brane in the $x^6$ direction. There are also $N_f/2$
D6--branes between the two half--fivebranes, along the $x^6$ direction.
Moving the NS half--fivebrane through the $N_f/2$ D6--branes generates
$N_f/2$ additional D4--brane connections between the D6--branes and
the NS half--fivebrane. Of course, we would expect this to be still true, as
locally nothing significantly new is happening which would lead one to
expect that new phenomena are occurring to change the massless
spectra. We therefore can draw a figure much like Fig.~2, the only
modification being the addition of the mirror plane. (See Fig.~5.)

\midinsert{
\vskip0.1cm
\hskip3.1cm
\epsfxsize=3.5truein\epsfbox{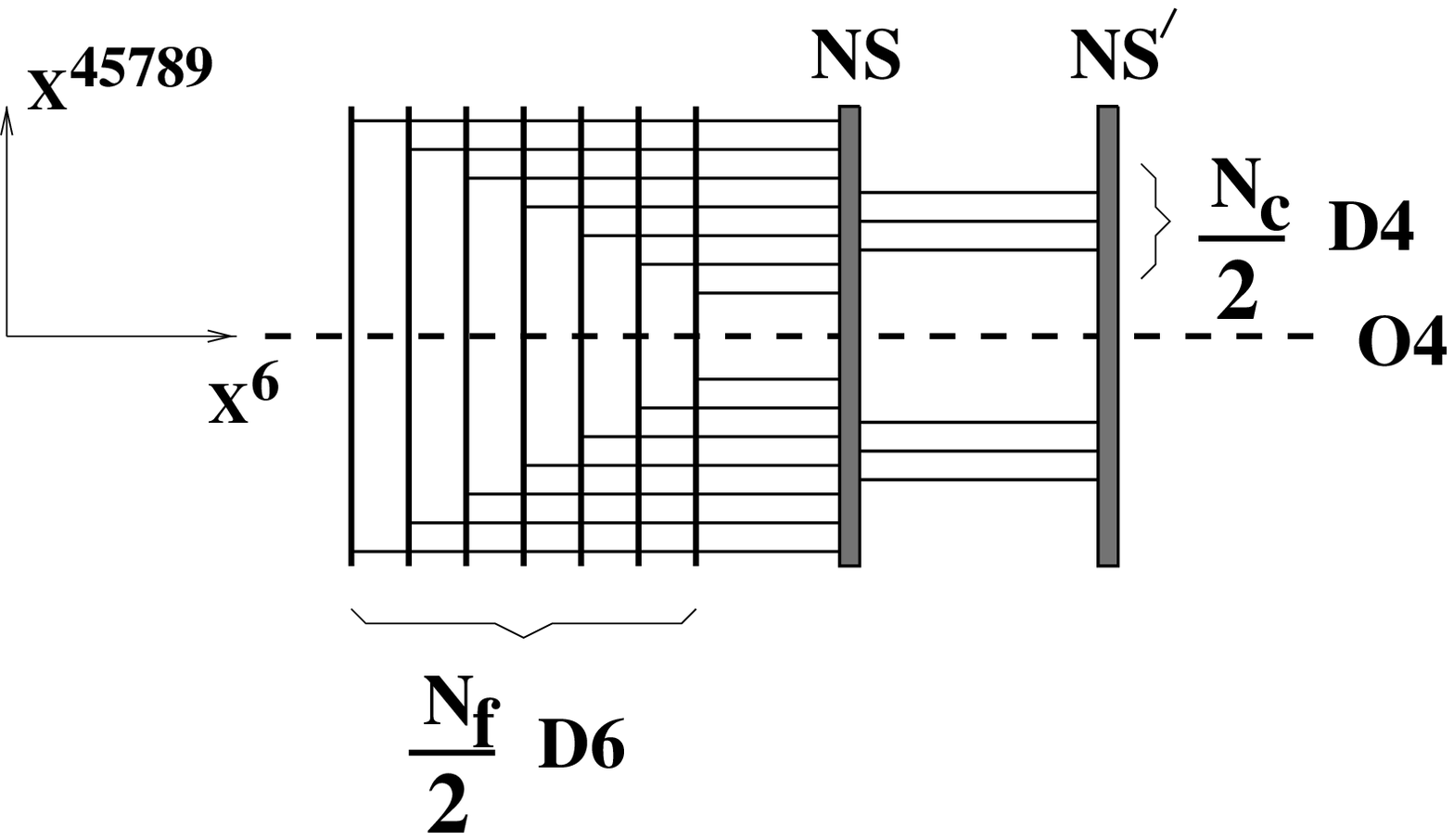}
\vskip0.1cm
\centerline{\sl Figure 5.}}\endinsert

The massless spectrum is now realised entirely by 4--4 strings, either
stretching amongst the $N_c/2$ D4--branes to give the gauge sector,
between the $N_f/2$  and the $N_c/2$ D4--branes to give $N_f$ massless
quarks, or amongst the $N_f/2$ to give flavour symmetry.
 
It is interesting to note at this stage that something non--trivial
has happened to the O4--plane. All of the possible non--Abelian
symmetries --- both gauge and flavour --- are carried by D4--branes,
in contrast to the earlier situation where the flavour sector was
carried by D6--branes. We have been careful to ensure that we have
done nothing to the spectrum, so this is the same model as we had
before moving the half--fivebrane past the D6--branes. However, we noted
previously that the D6--branes carried a $USp(N_f)$ symmetry whenever
the D4--branes carried an $SO(N_c)$ symmetry and {\it vice versa}. So
in this new situation, it must be that the D4--branes to the {\sl
left} of the NS half--fivebrane carry the same non--Abelian symmetry as the
D6--branes. Recall that the difference between the $SO$ or $USp$
choice was correlated with the sign of $\Omega^2$. Recall also that
the sign of the $A^{(5)}$ charge of the orientifold was correlated
with the sign of $\Omega^2$. Upon examination of Fig.~5., we are
therefore able to conclude that, moving along $x^6$, the sign of the
$A^{(5)}$ charge of the O4--plane flips as it passes a half--fivebrane
(NS) and then (by symmetry) flips back again as it passes the
NS$^\prime$ brane. Along the $x^6$ directions, the half--fivebranes act
as `domain walls' with respect to the orientifold charge. They
themselves have opposite twisted sector charges. We will use these
observations to our advantage as we proceed.

Ultimately, we are going to have to approach the strong coupling
singularity where we move the half--fivebranes to the same $x^6$
positions. Notice that this is strong coupling for both the field
theory (whose coupling goes inversely with their $x^6$ separation) and
for the string theory (because the dilaton blows up at the fivebrane
cores), as it should be.  This is the only place where a new
phenomenon can occur, and it happens just at the point where our
ignorance about how to compute is greatest.

Let us assume for a moment that we have passed the NS half--fivebrane
through the NS$^\prime$ brane successfully, passing to the other side,
and recovered a candidate for the `magnetic' dual theory. Let us see
what we can say about this new configuration. Taking what we have
learned from the $U(N_c)$ situation with $N_f$ flavours, our first
guess might be that perhaps there are now $(N_f{-}N_c)/2$ D4--branes
between the half--fivebranes by analogy. Indeed this was justified in that
case by passing one brane around the other.

After the fact, one can see that there is another argument for that
resulting $N_f{-}N_c$ situation, based upon the fact that it is the
only assignment of connecting D4--branes which preserves the local
`linking number' assignments to the fivebranes, following the
arguments of ref.\hanany.  The linking number between two branes is a
topological invariant calculated by integrating one brane's potential
(for which it a source) over the worldvolume of the other brane. One
must also take into account the presence of the endpoints of other
branes which end on the world volumes of the two branes in question,
because the endpoints act as sources in the worldvolume theories. In
the case of the $U(N_c)$ situation, the fivebranes have simply
exchanged their positions on the $x^6$ line, producing no change in
the contribution to linking number which involves their properties as
sources, as they are identical. There is no option in preserving the
total linking number but to redistribute the endpoint sources by
reconnecting with $N_f{-}N_c$ D4--branes between the fivebranes.

It is not clear whether such a linking number assignment is able to
restrict the physics in this case. We cannot completely compute the
linking number in this situation as there is not enough knowledge
about the detailed twisted sector couplings of the half--fivebranes and
the orientifold. What we do know is that passing one half--fivebrane
though the other is {\sl not} completely analogous to the situation
reviewed above, involving whole fivebranes. Due to the subtlety we
noticed earlier concerning their role in flipping the sign of the
orientifold's charge as one moves along the $x^6$ direction, we know
that these are {\sl not} identical objects under exchange. They carry
opposite amounts of twisted sector charge.

We therefore conclude that we will not simply get $(N_f{-}N_c)/2$
D4--branes, which would be the case if we had passed identical objects
through each other, but $(N_f{-}N_c)/2{+}\alpha$, where $\alpha$ is to
be determined.  The value (including the sign) of $\alpha$ is
ultimately computable with more knowledge about the twisted sector
charges, which we do not have.

The result $\alpha{=}\pm2$ (for $SO(N_c)$ and $USp(N_c)$,
respectively) suggests itself, by comparison to the magnetic theory we
are trying to recover. We cannot independently justify it at this
stage of the discussion because we have no way of doing a strong
coupling calculation. In the next section we will justify the claim
that $\alpha{=}\pm2$.

Assuming the result $\alpha{=}\pm2$ for now, the final situation is
thus as follows: After moving the NS half--fivebrane from the left, through
the NS$^\prime$ brane to the right, we have a net number of
$(N_f{-}N_c)/2\pm2$ D4--branes stretched between the two half--fivebranes
which will contribute to the massless spectrum.  We have $N_f/2$
D6--branes to the far left, with one D4--brane each stretched between
them and the NS$^\prime$ brane. (See Fig.~6.)

\midinsert{
\vskip0.1cm
\hskip3.0cm
\epsfxsize=4.0truein\epsfbox{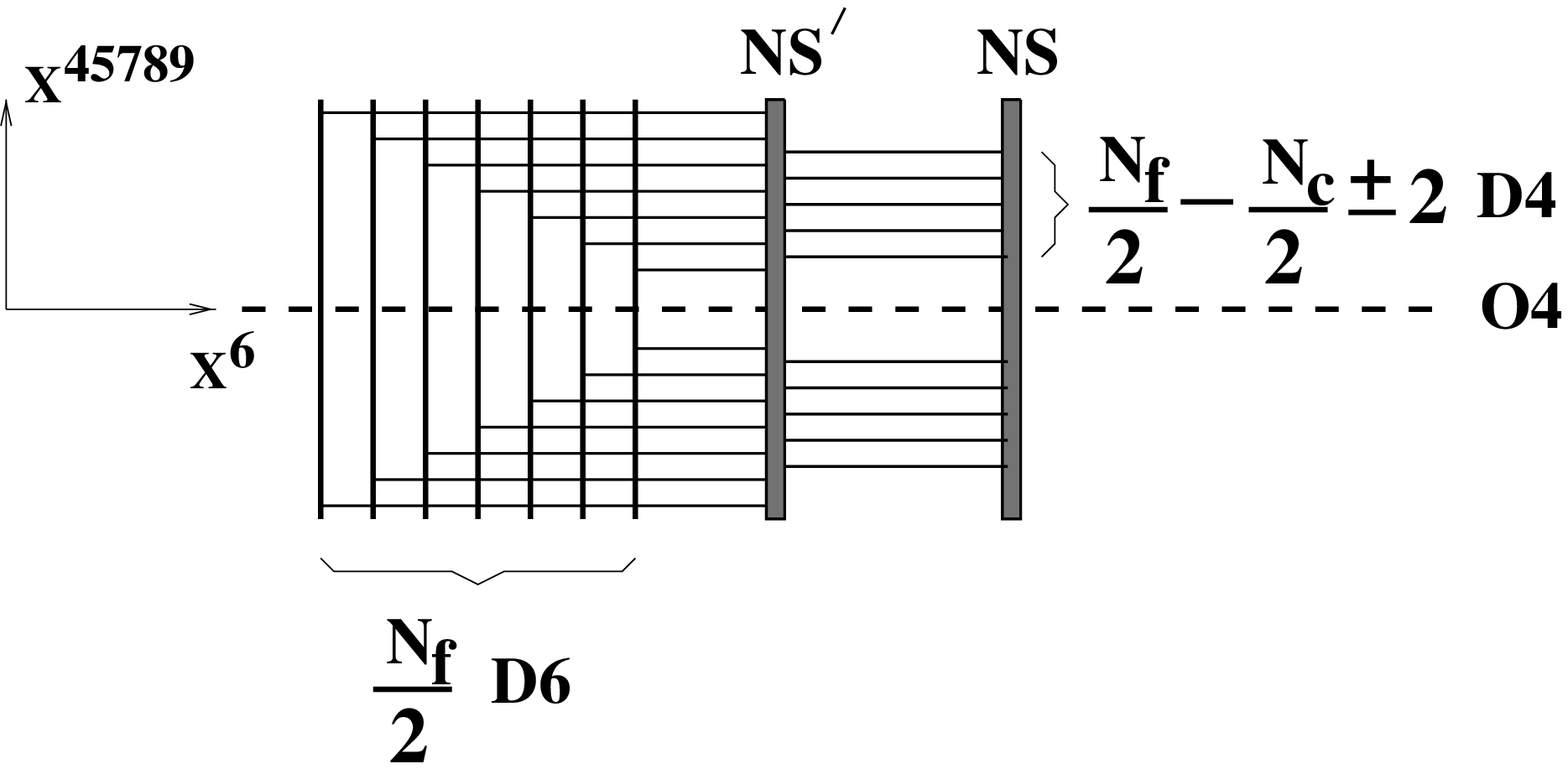}
\vskip0.1cm
\centerline{\sl Figure 6.}}
\endinsert

This gives an $SO/USp(N_f{-}N_c\pm4)$ gauge group coming from the 4--4
strings between the fivebranes, with 
$N_f$ quark flavours in the fundamental (coming from the
4--4 strings connecting the two different D4--brane families). There
is a meson associated to the $(x^8, x^9)$ fluctuations of the $N_f/2$
D4--branes.  As the D4--branes are precisely parallel to the
O4--plane, the meson is the (anti) symmetric part of $M_{ij}$ with
couplings to the quarks in the superpotential, as before.

So we see that our orientifolded weaving arrangement has reproduced
the dualities of refs.\refs{\seiberg,\intriligatorfour}.  In order to
recover this positive result, we had to guess that there was a
discrepancy of $\pm2$ physical D4--branes after we passed the
half--fivebranes through each other, over and above the appearance of
$(N_f-N_c)/2$ one might guess from trying to generalise
ref.\elitzur. { That there is a different number than $(N_f-N_c)/2$
was a justified assumption, due to the presence of the orientifold.}
However, fixing it to $(N_f-N_c)/2\pm2$ needs independent strong
coupling information about the string theory, which we discover in the
next section.

\newsec{\bf Strong coupling and $\N{=}2$ physics.}
Our goal is to find some independent means of deducing that there are
precisely {\sl two} new D4--branes which must appear (or disappear)
when the NS--NS half--fivebranes pass through each other from the electric
to the magnetic theory (or {\it vice versa}), in the presence of the
orientifold. We need some sort of clue about where these extra branes
might come from, and why we did not see them in the weak coupling
theory (at least in the massless spectrum).

The clue appears when we deform our brane configuration to one in
which the four--dimensional field theory has $\N{=}2$ supersymmetry.
Consider for a moment the same scenario which we had before, but with
the NS$^\prime$ half--fivebrane extended in the $(x^4,\,x^5)$
directions and pointlike in the $(x^8,\,x^9)$ directions, {\it i.e.,}
parallel to the NS half--fivebrane.  With the two half--fivebranes
parallel, there are now twice as many supersymmetries as in the
previous situation, and we have $\N{=}2$ supersymmetry in our field
theory. (As recently described in ref.\barbon, such a configuration is
continuously connected to the original situation by a rotation of the
NS$^\prime$ brane.)

Notice, however, that it is still impossible to move the half branes
around one another. (It is worth noting that now that the
half--fivebranes are oriented the same way, they can move together,
cancel their opposite twisted sector charges, and move off the
orientifold, together with their mirror partner moving the opposite
way. This is a degree of freedom which was not available in the
$\N{=}1$ situation.) We still have an unavoidable strong coupling
singularity.  There is some hope, though, that the exactly solved
$\N{=}2$ field theory might help us to understand aspects of the
strongly coupled dynamics of these branes --- specifically, the effect
of passing two NS half--fivebranes through each other\foot{Dualities
of $\N{=}2$ field theories have previously been used to understand
$\N{=}1$ dualities in
\APSei\ and \APS.}.

Let us focus on the case of $SO(N_c)$ with $N_f$ flavours and $N_c$
even. If we study our $\N{=}2$ theory on the Coulomb branch, we have
gauge symmetry $U(1)^{N_c/2}$, with $N_f$ charged hypermultiplet
fields. In this phase of the theory, much is known about its exact
structure\seibergwitten. In particular, the Coulomb branch of the
theory is controlled by the properties of an associated Riemann
surface \AS:

\eqn\polyone{y^{2}=x\left(\prod_{a=1}^{N_c/2}(x-\phi_a^2)^2
 %       +\Lambda^{2N_c-2N_f-4}
+x^2\prod_{i=1}^{N_f}(x-m_i^2)\right).}

Here, $m_i$ are the hypermultiplet mass parameters, $\phi_a$ are the
vev's of photons in the theory with generic gauge group $U(1)^{N_c/2}$,
and we have set the QCD scale $\Lambda$ equal to 1. This equation
describes a genus $N_c/2$ Riemann surface $\Sigma$ 
as a double sheeted plane. The space of possible vacua of the Coulomb
phase is the moduli space of such curves, and the prepotential for the
low--energy field theory is determined by the periods and residues
of an associated one--form $\lambda$ on $\Sigma$ \seibergwitten:

\eqn\LAMBDA{
        \lambda = {\sqrt x\over 2\pi i}d\,\log\left( 
        {x\prod(x-\phi_a^2) - \sqrt x y \over 
         x\prod(x-\phi_a^2) + \sqrt x y} \right) 
                }

The residues of $\lambda$ are linear combinations of the quark masses
$m_i$, and are located at the points $x=m_i$.  It is important to note
that there is no residue or monodromy around the point $x{=}0$, where
there appears to be interesting behaviour due to the $x^4$ term in the
curve \BL.  This point will be of great
interest to us below.

In the $\N{=}2$ theory, realised by D--branes, we can move through the
Coulomb branch by simply moving the D4--branes around, giving them
arbitrary and independent positions on the $(x^4,x^5)$ plane. The
strings which were connected to coincident D4--branes to give the
non--Abelian gauge symmetry are now massive, their masses being
proportional to the distances (in the $(x^4,x^5)$ directions) between
the various branes.

{\sl We propose that an identification should be made between the
abstract cut plane describing the $\N{=}2$ field theory's vacua in the
Coulomb phase and the $(x^4,x^5)$ part of the world--volume of the
NS--NS half--fivebranes, where the D4--branes end.}  Indeed, there is a
one--to--one correspondence between the masses and vevs parameterizing
the Coulomb branch and the D4--brane positions.

The precise correspondence is most easily made using a different
parameterisation of the curve \polyone. Substituting $x{=}z^2$ and $y{=}z
w$ (a generalization of the isogeny transformation of \DS\ and \AS)
gives the genus $N_c$ curve

\eqn\polytwo{w^2=\prod_{a=1}^{N_c/2}(z^2-\phi_a^2)^2
 %       +\Lambda^{2N_c-2N_f-4}
+z^4\prod_{i=1}^{N_f}(z^2-m_i^2).}

This curve $\widetilde\Sigma$ is a double cover of \polyone, with the
projection identifying $z$ with $-z$.  The previous solution is
obtained after modding out by this identification, leaving a set of
periods and one--forms in one--to--one correspondence with those of
the curve
\polyone\ \DS. 
 
In the new parameterization, we expect that an identification of the
form $z{=}x^4{+}ix^5$ may be made.  Then the above curve \polytwo\
should be viewed as being embedded in the covering space of the
orientifold. The two sheets of $\widetilde\Sigma$ are to be identified
with  the $(x^4{+}ix^5)$--planes of the two half--fivebranes. Punctures of
the curve are to be identified with D4--branes ending on the NS
half--fivebranes at the corresponding locations.  Reading off from the curve
and its associated one--form, we see that for every D4--brane ending
at a position $z{=}m_i$, there is another brane at $z{=}-m_i$. In the
weak--coupling limit of small $\Lambda$, there are also $N_c$ paired
D4--branes located at $x{=}\phi_a$ and $x{=}-\phi_a$. On the other
hand, the original curve \polyone\ embeds naturally in the
orientifold, and describes $N_c/2$ D4--branes at $x{=}\phi_a^2$. The
overall factor of $x$ indicates the presence of the orientifold plane
at $x{=}0$.

The next thing to do is to try to understand some of the features of
the string theory which might be immediately learned from this
correspondence. First of all, the QCD scale $\Lambda$ (which we have
set to one in Eqs.\polyone\ and \polytwo) characterizes the widths of
the cuts in the $z$--plane. These cuts can be thought of as tubes or
handles connecting the two sheets of the Riemann surface. At finite
$\Lambda$, the embedding we have described should thus be modified by
adding a compactified dimension to the D4--branes that run between the
half--fivebranes.  In this way, the solution of the $\N{=}2$ field theory
reveals the internal structure of D4--branes as objects of thickness
$\Lambda$, which connect smoothly to the NS half--fivebranes according to
the geometry of the Seiberg--Witten curve.  At strong coupling,
$\Lambda$ becomes large (in an asymptotically free theory) and the
internal structure becomes more apparent.\foot{Ultimately, given that
the strong coupling limit of the theory (which is locally type~IIA) is
supposed to be M--theory, we expect that the NS--NS fivebranes and
D4--branes all become M5--branes in different configurations in an
eleven dimensional theory. (The D4--branes unwrap a hidden leg wrapped
around the hidden eleventh direction.) Alternatively, if we had first
done a $T_{23}$--duality (the D4--branes becoming D2--branes and the
D6--branes becoming D4--branes) and then gone to strong coupling,
there would be an interesting M--theory configuration involving two
M5--branes with stretched membranes between them. The strong coupling
singularity of the string theory where they coincide is then
identified with a point at which tensionless strings could arise.}

If we are to identify all of the cuts and punctures of the $z$--plane
with the locations of D4--branes ending on the NS half--fivebrane, we should
also interpret the $z^4$ factor in the second term of the polynomial
in
\polytwo. In the expression \polyone, written in terms of the `physical'
variables, where mirror points are removed,  it is clear that  {\sl this
point should be identified with two extra  D4--branes}, which are forced
to live on the orientifold, at $z=0$. We should be careful though, as
the introduction of two extra D4--branes should naively change the
physics even away from strong coupling. 

Recall that there is no non--trivial physics (associated with stable
states) to be found in the $\N{=}2$ theory by examining monodromies
around this point $z{=}0$. \BL\  Correspondingly, there should be no new
physics arising in the weak coupling string theory either.  This must
mean that generically there are no new massless states coming from
fundamental strings stretching from these branes to any other branes
in the theory.

As far as the weakly coupled massless spectrum of the string theory is
concerned, these two extra branes must remain completely invisible 
throughout our discussion of the previous section, until we come to the
strong coupling regime. There, we anticipated that some extra branes
appear in the theory which stay in the spectrum as we move to the
magnetic theory. We had no means of fixing the number of such branes.
{\sl It is our conjecture that these `hidden' D4--branes, apparent in
the $\N{=}2$ theory's polynomial, are exactly the two D4--branes which
we sought in the previous section.} At the
point where the NS--NS branes become coincident in the $x^6$ direction,
these two branes appear in the theory on the same footing as all of the
other branes, contributing to the massless spectrum as we move off to
the magnetic theory. 

We also expect that precisely the reverse must happen upon moving from
the magnetic to the electric theory. This is perfectly consistent with
the fact that neither NS--NS half--fivebrane can move off the
orientifold and circle the other, a procedure that would result in an
unacceptable increase in the number of new branes.

Similar phenomena, of new states suddenly appearing in regions of
moduli space, occur when one crosses so--called surfaces of marginal
stability \seibergwitten\ in the moduli space of $\N{=}2$ gauge
theories. As one crosses such surfaces (typically of codimension one),
certain BPS states which were stable before crossing become unstable
and {\it vice versa}. In the case at hand, we may suppose that the
string states we might normally associate with the pair of D4--branes
stuck on the orientifold are unstable to decay when the distance
between the half--fivebranes is finite, so that their presence does not
directly influence the spectrum. Such a state would consist of a
string connecting a `hidden' D4--brane at $z{=}0$ to a gauge D4--brane
at $z{=}\phi_a$, together with its mirror image, running from $z{=}0$
to the mirror D4--brane at $z{=}{-}\phi_a$. We can imagine that this
state would be unstable to decay to a state running directly between
the other D4--brane and its mirror, which is already in the spectrum.
Once the half--fivebranes cross and the extra D4--branes at $z{=}0$ become
part of the magnetic gauge configuration, the string states would
become stable gauge bosons.
 
This completes our justification for picking $\alpha{=}2$ for the
$SO(N_c)$ theory in the previous section. As we can continuously move
from our $\N{=}1$ configuration to the $\N{=}2$ situation by   rotating
a fivebrane\barbon, and from there move smoothly to the Coulomb phase
where we see a sign of the two extra D4--branes, we expect that we
should take their presence seriously, and anticipate that they might be
relevant in the $\N{=}1$ theory. Admittedly, given the extra
supersymmetry and the other special features of the rotated theory, we
do not expect to be able to infer too much about the $\N{=}1$ theory
this way, but we expect that at least the number of these hidden branes
is preserved under the route we just described. In addition, we expect
that given the similarities of the $USp(N_c)$ theory to the $SO(N_c)$
theory from the point of view of string theory, the {\sl disappearance}
of two D4--branes as one goes to the magnetic theory is also plausible.

\newsec{Adjoint Matter}

In this section, we briefly present our speculations (based on the
 conjecture of ref.\elitzur) on how we expect the inclusion of adjoint
 matter into our orientifolded models to work.

The models discussed so far are in fact closely related to models with
a single matter field transforming in the adjoint of the gauge
theory. As shown in \barbon, rotation of either the NS$^\prime$
fivebrane or the NS fivebrane into the $(x^4, x^5)$ or $(x^8, x^9)$
directions preserves supersymmetry and when the two half--fivebranes
lie in the same orientation, $\N{=}2$ supersymmetry is restored. The
$N_c/2$ D4--branes are then free to move in the two of these
directions that are shared by the world volumes of the fivebranes and
correspond to the adjoint matter field's vev. Thus, in the $\N{=}1$
configuration, the relative rotations of the fivebranes corresponds to
a mass term for the adjoint fields.

As proposed in ref.\elitzur, different superpotential terms for the
adjoint may be included by the placement of extra NS fivebranes. In
the $SO/USp$ case the first non--trivial case is when we include one
extra {\sl whole} NS--NS fivebrane\foot{Adding an odd number of
half--fivebranes is probably not consistent. This would only flip the
O4--plane charge an odd number of times.}, coincident with the NS
half--fivebrane, but oriented like the NS$^\prime$ brane. We expect
that the resulting superpotential is $\tr X^4$.

In general the addition of $k$ extra coincident NS--NS fivebranes
(oriented like the NS$^\prime$ half--fivebrane) should generate the
superpotential term $\tr X^{2(k+1)}$.  The dualities of
\intriligatorfive\ follow simply from this construction and our
previous deductions. Each of the $2k{+}1$ half--fivebranes may be
moved through the D6--branes, creating $N_f/2$ D4--brane connections
to those D6--branes. Whole fivebranes may now freely be moved
through the NS$^\prime$ half--fivebrane carrying their $N_f$ connections
to the D6--branes. As these are whole fivebranes, we do not expect any
new physics (over and above that found in ref.\elitzur) when we move
them through. (In fact, as mentioned earlier, we expect that whole
fivebranes can move off the O4--plane.)

Finally we move the NS half--fivebrane through the NS$^\prime$
half--fivebranes with the resulting configurations discussed in the
previous two sections.  The final `magnetic' configuration has the NS
and NS$^\prime$ fivebranes interchanged and connected by $kN_f{+}(N_f
{-}N_c)/2 \pm 2$ D4--branes corresponding to the $SO/USp((2k{+}1)N_f
{-}N_c\pm 4)$ dual gauge symmetry. The dual theory also possesses an
adjoint field $Y$.  There are $N_f(2k{+}1)/2$ D4--branes connecting the
NS$^\prime$ and $N_f$ D6--branes.  4--4 strings in the final
configuration supply $N_f$ dual quarks.
 
The $N_f^2(2k{+}1)/2$ connections between D4 and D6--branes which are
free to move in the $(x^8,x^9)$ directions correspond to $(2k{+}1)$
mesons $M_n$ in the $N_f(N_f{+}1)/2$ and the $N_f(N_f{-}1)/2$
representations of the flavour group. The resulting superpotential is
of the form
\eqn\final{\tr Y^{2(k+1)} + \sum_{n=0}^{2k} M^{rs}_n q_r Y^{2k-n} 
{\tilde q}_s}
where $r$ and $s$ are flavour indices.

\newsec{Closing Remarks}

This paper presents a framework in which the electric/magnetic dual
descriptions of the physics of $\N{=}1$ supersymmetric $SO/USp(N_c)$
gauge theories with $N_f$ quarks are embedded into string theory
configurations.  The idea is to realise the duality as a consequence of
geometrical rearrangements of configurations of extended objects in
string theory, in the spirit of refs.\refs{\hanany,\elitzur}.

The addition of an orientifold plane allowed us to describe the
$SO/USp(N_c)$ gauge theories as a modification of the presentation of
ref.\elitzur\ for the $U(N_c)$ gauge theories.  Although this is a
simple modification to perform, it has very crucial consequences. It
forced us go through a non--trivial situation (passing half--fivebranes
through each other) in going between the magnetic and electric
descriptions. Such a singular situation could be avoided in the case
of $U(N_c)$, as shown in ref.\elitzur, but not here.

We emphasize again that the necessity of going through such a
singular configuration is not merely an inconvenience of the
orientifold description. It is string theory's way of reproducing
physics which is already anticipated from the point of view of field
theory. In particular:
 
\item\item{\it (i)} Due to the absence of
a $U(1)$ center for the $SO/USp$ gauge groups, there are no allowed
Fayet--Iliopoulos terms which may be included corresponding to the
freedom to perform a movement which avoids the singularity.

\item\item{\it (ii)} There {\sl must be} a phase transition in 
going between the electric and magnetic descriptions of $SO(N_c)$
gauge theories with quarks. That this is not the case for $U(N_c)$ is
signaled by the possibility of avoiding the singular
configuration. The only essential difference encountered between
performing the rearrangement of branes for $U(N_c)$ and for the cases
studied here is the necessity of the singular configuration and so we
expect that this is related to the presence of the phase transition.

Due to the lack of a description of orientifolded NS--NS fivebranes,
we were in an even worse position to describe the singular situation
when they overlap than the analogous case for $U(N_c)$. However, we
were able to smoothly deform the theory to an $\N{=}2$ model in its
Coulomb phase, where we were able to check some of our assumptions
about the new features which must arise at point where the NS--NS
fivebranes are coincident. To do so, we were able to 
identify the auxiliary higher genus (Seiberg--Witten) surface
associated with the vacua of the $\N{=}2$ theory with the
configurations of fivebranes and D4--branes which was present. This
identification allowed us to identify a pair of extra D4--branes which
must appear in the theory as we go between phases\foot{One can also
carry out this procedure in the $U(N_c)$ case, where, upon examination
of the Seiberg--Witten curve, it is clear that there are no extra
D4--branes.}.

In closing, we note that there are many avenues of investigation to
pursue further. Chief among these is the issue of how much more about
the physics of extended objects (and the resulting field theories they
encode) can one learn by studying the powerful results of $\N{=}2$
field theory.  In this paper, we have found a correspondence between
the Seiberg--Witten curve and configurations of D4--branes ending on
NS--NS fivebranes.  There are undoubtedly many more entries to be put
into the dictionary which translates between the physics of extended
objects and the physics of exactly solvable field theories, which all
will be of great value in continuing to understand duality in both
field theory and string theory.

%%%%%%%%%%%%%%%%%%%%%%%%%%%%%%%%%%%%%%%%%%%%%%%%%%%%%%%%%%%%%%%%%%%%%%%%%%
\bigskip
\medskip
%\vskip0.5truecm

\bigskip
\bigskip
\noindent
{\bf Acknowledgments:}

\noindent
We would like to thank Philip Argyres and Michael Crescimanno for
discussions, and Eric Gimon for helpful comments on the manuscript.
  The work of NE was supported in part by the Department
of Energy under contract
\#DE--FG02--91ER40676.  ADS's research was supported in part by DOE
EPSCoR grant \#DE--FC02--91ER75661 and by an Alfred P. Sloan
Fellowship.
%CVJ was supported by clean living and good music.
NE would also like to thank the Department of Physics and Astronomy at
the University of Kentucky for hospitality while some of this research
was carried out.  CVJ would like to thank the Physics Department at
Stanford University for their kind hospitality and enjoyable working
environment while this research was completed.
\listrefs 
\bye